\documentclass[a4paper,11pt]{article}
\usepackage{jheppub_kim}
\usepackage{pdflscape}
\usepackage{amsmath}
\usepackage{amssymb}
\usepackage{dcolumn}
\usepackage{bm}
\usepackage{color}
\usepackage{epsfig}
\usepackage{amsfonts}
\usepackage{graphicx}
\usepackage{subfigure}
\usepackage{dcolumn}
\begin{document}
\title{The investigation of  quark - antiquark  potential in plasma  with hyperscaling violation background}
\author{S. Tahery,}
\author{J. Sadeghi}
\affiliation{Sciences Faculty, Department of Physics, University of Mazandaran, 47416-95447, Babolsar, Iran}
\emailAdd {s.tahery@stu.umz.ac.ir}
\emailAdd {pouriya@ipm.ir}
\abstract{In this paper,  we  investigate  imaginary part of the potential for a moving quarkonia in plasma. In order to verify the  validity of imaginary part of the potential we employ  hyperscaling violation metric background. In QCD system the imaginary part of potential shows the decay behaviour and the corresponding  potential should be negative.  The corresponding  potential always appears in QCD as an exponential function. We find general constraints for Im$V_{Q \bar{Q}}$ at the near horizon limit and apply them for some known cases. Finally, we find the suitable  spatial dimension $d$, $z$ and $\theta$  also rapidity $\eta$ for two interesting cases. Our calculation with AdS/CFT stuff guarantees the dynamical parameter as a $\theta$ in hyperscaling violation metric background.  We also present the  hyperscaling violationg metric in this context  very close to YM theory with finite temperature.}
\maketitle
\section{Introduction}
The AdS/CFT  is a correspondence \cite{jmm,ssg,ew,oas,kli,bim} between a string theory in AdS
space and a conformal field theory in the physical space-time. It leads
to an analytic semi-classical model for strongly coupled QCD. It has
scale invariance, dimensional counting at the short distances and color
confinement at  large distances. This theory describes the
phenomenology of hadronic properties and demonstrate their ability
to incorporate such essential properties of QCD as the confinement and
 the chiral symmetry breaking \cite{darl,otpt,crbe,hoba}.\\
When we want to study the $Q \bar{Q}$ interaction,  we should consider the effect of the medium in  motion of $Q  \bar{Q}$. Also we note that the interaction in this pair is not produced at rest in QGP, so  the velocity of the pair through the plasma has some effects on process of  the it's interactions. In that case, we account such effect to the corresponding system.
At finite temperature the interaction energy has a finite imaginary part.  It can be used for the estimating  thermal width of the quarkonia \cite{nbma,ybm}. As we know all researcher   before AdS/CFT method used   pQCD \cite{mlop} and lattice QCD  \cite{arth,gaca,gcs} approaches for the calculations of the Im$V_{Q \bar{Q}}$ correspond  to QCD in case of static $Q \bar{Q}$ pairs.\\
Here we discuss about gravity dual of QCD. Can we have an exact gravity dual of QCD? One can close to it to some extent \cite{astp}. What people do, is to start by considering $N=4$ SYM theory, which at long distances the theory reduces to pure Yang-Mills theory. But there are some differences between this YM theory and the gauge theory dual to the original  $N=4$ theory. Although the vacua of QCD and $N=4$ SYM theory have very different properties $N=4$ SYM at $T\neq 0$ with QCD $T> T_{c}$ (temperature $T_{c}$ of the crossover frome a hadron gas to quark-gluon plasma), many of the qualitative distinctions disappear or become unimportant. In particular, QCD above $T_{c}$ is not confining any more because its $T=0$ quasiparticles are hadrons within which quarks are confined is not relevant  above  $T> T_{c}$. Chiral symmetry is broken in QCD, but in QCD  $T> T_{c}$, the chiral condensate melt away and this distinctions between these two theories also vanishes. QCD is not scale invariant in spite of $N=4$ SYM, but  QCD above  (but not asymptotically far above) its $T_{c}$ is much more similar to $N=4$ SYM theory at $T\neq 0$ than the vacua of the two theories are, because at higher temperature than $T_{c}\sim 170$ Mev the quark gluon plasma becomes more and more scale invariant, at least in its thermodynamics \cite{adsqcd}. Although supersymmetry is explicitly broken at  $T\neq 0$ of $N=4$ SYM, but it does not play a major role in the caracterization of properties of the $N=4$ SYM plasma at nonzero temperature. In addition, QCD is an asymptotically free theory and high energy processes are weakly coupled at T above $T_{c}$ that is accessible to heavy ion collision experiments the QCD plasma is strongly coupled. For all these reasons and some others, the strongly coupled plasma of $N=4$ SYM theory has been studied by many people for understanding dynamics of deconfined QCD plasma. In an important work Chu and Matsui studied dynamical debay screening and inter-quark potential in a moving medium \cite{ddsc}.\\
The investigation of moving heavy quarkonia in a space-time  plays an
important role in the interaction energy \cite{mst,msd,mmk,gac} by AdS/QCD approach. The non-relativistic bound states in a moving thermal bath has been studied by \cite{nbs}. On the other hand   the heavy Quarkonium moving in a Quark-Gluon Plasma is another interesting work  which has been studied by \cite{drst}. Here we note that the different metric backgrounds lead us to face various effects of interaction energy.
The evaluation of Im$V_{Q\bar{Q}}$ will yield to determine the suppression of ${Q\bar{Q}}$ in heavy ion collision \cite{sif}. On the other hand the effects of deformation parameter in the thermal width of moving quarkonia in plasma has been studied in \cite{tec}.  Also the imaginary part of the static potential has been studied  as a observable and moving quarkonia  in strongly coupled anisotropic plasma \cite{ipap,osca,ihqm}. Here, in order to have Re $V_{Q\bar{Q}}$ and Im $V_{Q\bar{Q}}$  for ${Q\bar{Q}}$ in a plasma we use  boosted frame \cite{fn}.
From bottom-up  and top-down approaches and some gauge gravity dualities one can study the behaviour of some parameters in the corresponding theory. By bottom-up approach we consider different metric backgrounds and  study the behavior of some parameters  in the thermal width of a moving quarkonia in plasma. One interesting case is considering metrics which are dual to the field theories and  are scale invariant but not conformally invariant. As we know such corresponding metrics have a Lifshitz  (with hyperscaling violation) scaling symmetries at quantum critical points and a dynamical critical exponent called $z$. We note that  the behavior of the system with  hyperscaling violation metric backgrounds near to  phase transition, is characterized by the critical exponent $z$. Also we know that the  time and space in such background will be scaled differently. So the corresponding  metric is not invariant under the mentioned scaling. Here $d-\theta$ plays the role of an effective  dimensional space in the dual field theory  \cite{pd,sr},
and  theories with hyperscaling violation are intrinsically non-relativistic. From  bottom-up approach point of view we try to use such corresponding backgrounds as toy models for the meson system. All above informations give us motivation to study the validity of the imaginary  part of potential for a moving quark- antiquark in plasma media. So, here we take the hyperscaling violation metric background and derive the    Im $V_{Q\bar{Q}}$  . The negative condition of potential as    Im $V_{Q\bar{Q}}\leq 0$ leads us to arrange    $z$ , $\theta$ and $d$ with appropriate values. So, we organize the corresponding paper as follows. In section 2 we introduce the hyperscaling violation metric background. In that case, we review this metric from several point of view. Also, we investigate the relation between   $z$ , $\theta$ and $d$ which are investigated by several papers \cite{tsc,hybb,jqp}. In section 3 we have a review on some interesting hot QCD  works which discussed on decay width. In section 4 we employ the corresponding metric background.   Also, we resolve the imaginary part of the potential and show   Im $V_{Q\bar{Q}}\leq 0$ in the near horizon limit . In this section our attention is more to case of moving pair perpendicular to the joining axes of dipole.  In section 5, we extend our calculation to the general orientation of  ${Q\bar{Q}}$ at an arbitrary angle. In here also we show the validity of the Im $V_{Q\bar{Q}}\leq 0$ which is an important condition in QCD phenomena.  In section 6 we have conclusion and some results and suggestions for the future.

\section {Review of hyperscaling violation metric background}
 Before to discuss and review  of hyperscaling violation metric background, we need to explain about the modification and application of the metric in QCD. Here, we note that such metric is breaking  conformal invariance.
As was shown by Polchinski and Stassler \cite{jpol}, it is possible to simulate confinement introducing a cut off in the holographic
coordinate $z$. This kind of model is known as hard wall . In
hard-wall models, the space is confined to values $z \leq z0$, where
$z_0\approx \frac{1}{\Lambda}$ QCD (infrared cut-off) breaks conformal invariance, approving
the introduction of the QCD mass scale. Although phenomenologically they have problems, because
the obtained spectra does not have Regge behavior. To eliminate
this problem, it is necessary to introduce a soft cut off. The
soft-wall model is constrained to a z-dependent dilation field to
break conformal symmetry and submit confinement.  Here, without using the hard and soft-wall
method for  eliminating the above problems we employ the hyperscaling violation metric background. For this reason, 
in this section  we are going to review  of  hyperscaling violation (HSV) metric background.  Such a metric background is scale invariant but not conformally invariant. In order to study  corresponding background first we consider the following metric,
\begin{equation}\label{metric1}
ds^2=-\frac{1}{r^{2z}} dt^2+ \frac{1}{r^2}(dr^2+dx_{i}^{2}),
\end{equation}
which is invariant under scaling
 $t\longrightarrow\lambda^{z} t $ , \quad $x_{i}\longrightarrow\lambda x_{i} $, and \quad $ r\longrightarrow\lambda r$.
 These metrics are exact solutions to gravitational theories coupled to a appropriate matter,
with an abelian gauge field in the bulk. By including an abelian gauge field and scalar
dilaton, one can construct the full class of metrics \cite{ppc,hoc,ptb,hcd,eff,dwh,tbc,hddb,hfn,cdab,hfs,hfsi,heef}, so which is given by,
\begin{equation}\label{metric2}
ds^2_{d+2}=r^{-2\frac{d-\theta}{d}} (-r^{-2(z-1)} dt^2+dr^2+dx_{i}^2),
\end{equation}
where $z$ and  $\theta$  are  dynamical critical exponent of hyperscaling violation metric  \cite{scsd}. As mentioned before, this metric is not scale invariant under above scaling.
So, if  we consider  some finite temperature in this theory we have to account some $f(r)$  in  the corresponding metric which is given by following equation,
\begin{equation}\label{metric3}
ds^{2}_{d+2}=e^{2A(r)} (-e^{2B(r)} f(r) dt^2+ \frac{dr^2}{f(r)}+dx_{i}^2).
\end{equation}
Then the temperature is proportional to a power of $r_{h}$ and for general $A(r)$ and $B(r)$ the temperature also depends on them. Moreover, in the gravity side we should take $r_{F} <r_{h}$ and $r_F$ is the inverse scale of the Fermi surface. In order to have black hole solution we have to consider following background \cite{hfnf},
\begin{equation}\label{metric4}
ds^{2}_{d+2}=\frac{R^2}{r^2} (\frac{r}{r_{F}})^{\frac{2\theta}{d}} (-r^{-2(z-1)}f(r) dt^2+\frac{dr^2}{f(r)}+dx_{i}^{2}),
\end{equation}
with $f(r)=1-(\frac{r}{r_{h}})^{d+z-\theta}$ , $r_{F}<r_{h}$ and the temperature will be as $T=\frac{1}{4\pi} \frac{\vert d+z-\theta \vert}{r^{z}_{h}} .$
Here we note that the  the Null Energy condition imposes $(d-\theta)(d(z-1)-\theta)\geq 0$, \quad \quad $(z-1)(d+z-\theta)\geq 0$ \cite{ucbe}.\\
For non-supersymmetric case of  the type IIB string theory  one can obtain by following equation \cite{isfh},
\begin{equation}\label{solution}
z=\frac{3+3\gamma}{3-\gamma} ,\quad \quad \theta=\frac{12}{3-\gamma},
\end{equation}
where $ -1\leq\gamma\leq1$.  As mentioned before the hyperscaling violation metric  background has been discussed several times from different point of view. The most important things here is how we can arrange the exact values for the $z$, $\theta$ and $d$.

Here also note  all  solutions in references have been resulted in the near horizon limit.  So, for this reason in next section we will find Im$V_{Q \bar{Q}}$ in the near horizon limit. Also, here we want to arrange the values of  $z$, $\theta$ and $d$ with the help of a condition on the corresponding potential which  plays an important role in QCD. It means that the Im$V_{Q \bar{Q}}$ leads us to specify the mentioned parameter in the paper.
\section{Review of some interesting hot QCD works}
In this section we will review on some QCD works that discuss on inter-quark potential in the moving medium. In heavy ion collision experiments at the LHC, hadrons containing at least one heavy quark, are reviewed \cite{newhad}. In this work heavy-quark related observables are becoming increasingly important. The authors pointed out that quarkonium remains a coherent quantum-mechanical bound state in a thermal medium, with only the potential that binds it together getting modified by Debye-screening\cite{qgpf}. Technically, it implies that a suitable defined static potential may develope an imaginary part or it can be developed by determining a real-time static potential, through a spectral analysis of an imaginary-time wilson loop. But in \cite{qgpf}, heavy ion collision has been considered from view point of on of the main experimental signatures of the formation of a strongly coupled quark-gluon plasma (QGP), which is melting of quarkonium systems, like $J/\psi$ and excited states in the medium. In this case the main mechanism responsible for this suppression is color screening. But other works like \cite{mlop,recrde} suggest a more important reason than screening which is the existence of an imaginary part of potential. In the former the authors derive a static potential for a heavy quark-antiquark pair propagating in Minkowski time at finite temperature. In result, there is a Debye-screened potential could be used in models that attempt to describe the melting of heavy quarkonium at high temperatures. In that work, a lower bound of temperature, strong gauge coupling dependent has been suggested for width. In the latter paper the relation between cross section, decay width and imaginary potential of heavy quarkonium in a quark-gluon plasma has been studied. In this paper perturbative computations and lattice studies suggest the exictence of an imaginary part of potential and some general aspects of effective field theories for heavy quarkonium in the medium have been reviewd. In addition the physical phenomena behind the imaginary part of potential has been discussed. They show the importance of the imaginary part of potential and decay width for dissociation.\\
Thermal decay widths have been studied in the effective field theory framework \cite{stf,hqwe}. In the first work, the authors proceeded by two mechanism according to the imaginary part of the gluon self energy induced by the Landau damping phenomenon, and the quark-antiquark color singlet to color octet thermal break up. Their results are  in agreement
with a  calculation of the static Wilson loop at finite temperature and they found new contributions to the potential, both real and imaginary, which is relevant to understand the onset of heavy quarkonium dissociation in a thermal medium. In the second one  the heavy quarkonium energy levels and decay widths in a quark gluon plasma, has been calculated, also the implications of the results
concerning heavy quarkonium suppression in heavy ion collisions have been discussed. In \cite{pNRQCD} two different mechanisms contribute to the decay width. The authors investigate the relation between the singlet-to-octet thermal break-up and the so-called gluo-dissociation, a mechanism for quarkonium dissociation widely used in phenomenological approaches. They have shown that, under the scale hierarchy, the leading contribution to the thermal width of a quarkonium $1S$ state may be written as
a convolution integral of the gluon distribution function and a cross section. The analytic estimate of the imaginary
part of the binding energy and the resultant decay width were studied in \cite{anis} where   the form of
the real-time, hard thermal loop resummed  propagator for static gluons in the presence of
an anisotropies and the consequences for quarkonium binding have been discussed. It has been predicted that
the propagator develops an imaginary part due to Landau damping at high temperature. In \cite{comhe} complex heavy-quark potential at finite temperature from lattice QCD has been considered. In that work they calculate for the first time the complex potential between a heavy quark and antiquark at finite temperature across the deconfinement transition in lattice QCD. The real and imaginary part of the potential at each separation distance r is obtained from the spectral function of the thermal Wilson loop.
The effect of the imaginary part of the potential on the thermal widths of the states in both isotropic and anisotropic plasmas have
been studied in \cite{qstate}. In this paper calculation of  quarkonium binding energies by using a realistic complex-valued potential for both an isotropic and anisotropic quark-gluon plasma has been studied, where  the disassociation temperatures of the ground and first excited states considering both the real and imaginary parts of the binding energy have been determined. In, \cite{surv}  readers consider two mechanisms of charmonium attenuation by considering final state interactions, regarding to the fact that in heavy ion collisions, most of quarkonium observed experimentally are moving through the medium with relativistic velocities. One conclude they have finite probability to survive, even at infinitely high temperature. The bottomium suppression in PbPb
collisions at LHC energies is studied in \cite{pbpb}, where the authors considered collisional damping of some states as described in
a complex potential, screening of the real part of the potential and the subsequent decay cascade. In \cite{teasoo} the thermal width of heavy quarkonia moving in quark gluon plasma has been studied. It is an interesting work where one can see the velocity dependence of the thermal width of heavy quarkonia traveling with respect to the quark gluon plasma. This is calculated up to the NLO in perturbative QCD. It concluded that at the LO, the width decreases with increasing speed, whereas at the NLO it increases with a magnitude approximately proportional to the expectation value of the relative velocity between the quarkonium and a parton
in thermal equilibrium. In addition it has been shown such an asymptotic behavior is due to the NLO dissociation cross section
converging to a nonvanishing value in the high energy limit. In the regime relevant for dissociation and for very large velocities the thermal width decreases with increasing velocity of quarkonium \cite{hqm}. In this work by means of effective field theory techniques,  the modifications of some properties of weakly coupled heavy quarkonium states propagating through a quark-gluon plasma  has been considered at temperatures much smaller than the heavy quark mass. Two different cases are considered, the first case relevant for moderate temperatures, and the second one relevant for studying the dissociation mechanism. In the first
case they determine the perturbative correction to the binding energy and to the decay width of states
with arbitrary angular momentum, finding that the width is a decreasing function of the velocity.
A different behavior characterizes the second kinematical case, where the width of s-wave states
becomes a non-monotonic function of the velocity, increasing at moderate velocities and decreasing
in the ultra-relativistic limit.

\section{The Validity of Im$V_{Q\bar{Q}}$ for perpendicular case of dipole}
As we know from  Wilson loop one can obtain the potential of quark-antiquark with rectangular loop. On the other hand  the expectation value of Wilson loop in large $N$ limit with the condition of string worldsheet helps us to arrange the corresponding potential. By choosing different metric background in Nambu-Goto action and mapping potential give some configuration to the quark system \cite{wipo,fn}. So, we first derive the  Im$V_{Q\bar{Q}}$ when the dipole moves with velocity perpendicular  to the wind and rapidity $\eta$ \cite{sif}.  Then, we will apply such a process for a hyperscaling violation metric and represent a derivation of relation for the imaginary part of potential . In that case we have to consider the effects of worldsheet fluctuations around the classical configuration $r_{c(x)}$, which is given by
\begin{equation}\label{flucr}
r(x) = r_c(x) \rightarrow r(x) = r_c(x) + \delta r (x),
\end{equation}
where $c$ is the deepest position of the string in the bulk.
Then, the fluctuations should be taken into account in the partition function, so one can arrive to the following equation,
\begin{equation}\label{parfu}
Z_{str} \sim \int \mathcal{D} \delta r(x) e^{i S_{NG} (r_c(x) + \delta r (x))},
\end{equation}
where there is an imaginary part of potential in the action.  We divide the interval region of $x$  into $2N$ points and also consider   $N\longrightarrow\infty$  at the end of calculation finally  we arrive at,
\begin{equation}\label{parfu2}
Z_{str} \sim \lim_{N\to \infty}\int d [\delta r(x_{-N})] \ldots d[ \delta r(x_{N})]  \exp{\left[ i \frac{\mathcal{T} \Delta x}{2 \pi \alpha'} \sum_j \sqrt{M(r_j) (r'_j)^2 + V(r_j)}\right]},
\end{equation}
where,
\begin{equation}\label{mr}
M(r)\equiv G_{00}G_{rr},
\end{equation}
\begin{equation}\label{vr}
V(r)\equiv G_{00}G_{xx},
\end{equation}
\begin{equation}\label{pr}
P(r)\equiv {G_{xx}}^2,
\end{equation}
\begin{equation}\label{nr}
N(r)\equiv G_{xx}G_{rr}.
\end{equation}
The thermal fluctuations are important around $r_c$ which means that  $x=0$, because of that  we should expand $r_c(x_j)$ around $x=0$ and keep only terms up to the second order,
\begin{equation}\label{expr}
r_c(x_j) \approx r_c + \frac{x_j^2}{2} r_c''(0),
\end{equation}
By considering small fluctuations,  finally we will have the following expression,
\begin{equation}\label{vrj}
V(r_j) \approx V_c + \delta r V'_c + r_c''(0) V'_c \frac{x_j^2}{2} + \frac{\delta r^2}{2} V''_r,
\end{equation}
where $V_c\equiv V(r_c)$  and $V'_c\equiv V'(r_c)$.
We use equations (\ref{expr}), (\ref{vrj}) and (\ref{parfu2}),   one can derive $S^{NG}_j$, $C_1$ and $C_2$ respectively,
\begin{equation}\label{sj}
S^{NG}_j = \frac{\mathcal{T} \Delta x}{2 \pi \alpha'} \sqrt{C_1 x_j^2 + C_2}
\end{equation}
\begin{equation}\label{c1}
C_1 = \frac{r_c''(0)}{2} \left[ 2 M_c r_c''(0) + V_r' \right]
\end{equation}
\begin{equation}\label{c2}
C_2 = V_r + \delta r V'_r + \frac{\delta r^2}{2} V''_r.
\end{equation}
In order to have  $ Im V_{Q\bar{Q}}\neq 0$  the function in the square root of (\ref{sj}) should be negative. So, we consider the $j$-th contribution to $r_{c}$, which is given by,
\begin{equation}\label{ij}
I_j \equiv \int\limits_{\delta r_{j min}}^{\delta r_{j max}} d(\delta r_j) \, \exp{\left[ i \frac{\mathcal{T} \Delta x}{2 \pi \alpha'} \sqrt{C_1 x_j^2 + C_2} \right]}.
\end{equation}
 The condition of $\delta r_{min}<\delta r<\delta r_{max}$  leads us to have $C_1 x_j^2 + C_2 <0$. It means that such an expression guarantees the  $ Im V_{Q\bar{Q}}\neq 0$ which is important condition in our calculation. The exponent has a stationary solution when we have following equation,
\begin{equation}\label{ddelta}
D(\delta r_j) \equiv C_1 x_j^2 + C_2(\delta r_j),
\end{equation}
and the extremal values is,
\begin{equation}\label{deltar}
\delta r = - \frac{V'_c}{V''_c}.
\end{equation}
Here  the $ D(\delta r_j)<0 \longrightarrow  -x_c<x_j<x_c$ leads us to have an imaginary part in the square root and   $x_c$  is given by following equation,
\begin{equation}\label{xc}
x_c = \sqrt{\frac{1}{C_1}\left[\frac{V'^2_c}{2V''_c} - V_c \right]}.
\end{equation}
If the square root in (\ref{xc}) is not not real, we should take $ x_c=0$. After all these  pointed conditions  and using $ D(-\frac{V'_c}{V''_c})$ in $ I_j$ one  can approximate $D(\delta r)$ , so we have following equation,
\begin{equation}\label{ijapprox}
I_j \sim \exp \left[ i \frac{\mathcal{T} \Delta x}{2 \pi \alpha'} \sqrt{C_1 x_j^2 + V_c - \frac{V'^2_c}{2V''_c}} \right].
\end{equation}
The total contribution to the imaginary part  with the continuum limit  will have the following form,
\begin{equation}\label{imv1}
\mathrm{Im} \, V_{Q\bar{Q}} = -\frac{1}{2\pi \alpha'} \int\limits_{|x|<x_c} dx \sqrt{-x^2 C_1 - V_c + \frac{V'^2_c}{2V''_c}}\,.
\end{equation}
Here the necessity of the negative sign of formula is  completely clear. But  we note that there are some points about the sign of $ImV$ in here. As we know, the imaginary potential is related to the decay of bound states in QCD. Thus, one should have positive shift in the energy level of the bound state. In other words, the imaginary part of the potential shifts the Bohr energy level where it is related to $-ImV$  \cite{ipap}.
Here, the interesting fact  is that one does not choose the sign inside the square root and it is fixed,  because the action must be real in a classical solution. If we use the opposite sign, then we would have an imaginary action even in a classical solution. In this direct,  since we consider fluctuations over the classical solutions, the action can develop an imaginary part. So, because the Nambo Goto action is truly negative in a Minkowiski signature, one can not choose the sign in the action.
And finally, after evaluating the integral one can achieve the expression for the imaginary part of the potential, which is given by,
\begin{equation}\label{imv2}
\mathrm{Im} \, V_{Q\bar{Q}} = -\frac{1}{2 \sqrt{2} \alpha'} \sqrt{\tilde{M}_c} \left[\frac{\tilde{V'}_c}{2\tilde{V''}_c}-\frac{\tilde{V}_c}{\tilde{V'}_c} \right].
\end{equation}
 where $\tilde{M}(r)$ and $\tilde{V}(r)$ are defined as,
\begin{eqnarray}\label{mtilda}
\tilde{M}(r)\equiv M(r)\cosh ^2 \eta -N(r)\sinh ^2 \eta\ ,
\end{eqnarray}
\begin{eqnarray}\label{vtilda}
\tilde{V}(r)\equiv V(r)\cosh ^2 \eta -P(r)\sinh ^2 \eta\ .
\end{eqnarray}
The most important point here is to consider suitable metric background which is  black hole solution of the action. Here the corresponding metric give all information of $M(r)$, $N(r)$, $V(r)$ and $P(r).$ In here, there are some points  that should be considered as well.  Generally we can say that our solution of action may be appear in two cases with trivial and nontrivial dilaton fileds. Both two cases will be considered with configuration  of string frame.  In nontrivial case we have some warp factor which depends on dialton field, this leads us to consider more terms in our calculation. This case is more complicated than trivial case, so here we just consider the first order of dilaton.  About the terms of dilaton some people neglect it at the first step as we see in \cite{ugkm}. Then, one can check that the integral on the action corresponds to the worldsheet with different topology. This means that, we can do the string interactions and go to a higher order in the string perturbation theory. But now, for the leading order calculations in genus, we do not need to bother with this term even if the geometry has a dynamical dilaton. Therefore, we will use equations (\ref{imv2}), (\ref{mtilda}) and (\ref{vtilda})  for  the corresponding metric. We continue the  calculation  with the help of metric background (\ref{metric4}) and equations (\ref{mtilda}), (\ref{vtilda}) so we have,\\

\begin{equation}\label{g00}
G_{00}=\frac{R^2}{r^2} (\frac{r}{r_F}) ^{\frac{2\theta}{d}} \frac{f(r)}{r^{2(z-1)}}
\end{equation}
\begin{equation}\label{gxx}
G_{xx}=\frac{R^2}{r^2} (\frac{r}{r_F}) ^{\frac{2\theta}{d}}
\end{equation}
\begin{equation}\label{grr}
G_{rr}=\frac{R^2}{r^2} (\frac{r}{r_F}) ^{\frac{2\theta}{d}} \frac{1}{f(r)}.
\end{equation}
After some algebratic calculations, we arrive at following results,
\begin{equation}\label{vtildahsv}
\tilde{V}(r)\equiv \frac{R^4}{r_F^{C+4}}[(r^A-\frac{r^B}{r^{B-A}_{h}})\cosh ^2 \eta- r^C\sinh ^2 \eta],
\end{equation}
where we define,
\begin{eqnarray}\label{a,b,c}
A&=&\frac{4\theta}{d}-2z-2,\nonumber\\
B&=&\frac{4\theta}{d}+d-z-\theta- 2,\nonumber\\
C&=&\frac{4\theta}{d}-4.
\end{eqnarray}
On the other hand ,  we use equation (\ref{mtilda}) and  obtain following equation,
\begin{equation}\label{mtildahsv}
\tilde{M}(r)\equiv \frac{R^4}{r_F^{C+4}}[r^A\cosh ^2 \eta- \frac{r_h^{B-A} r^C}{r_h^{B-A}-r^{B-A}}\sinh ^2 \eta].
\end{equation}
From  equation (\ref{vtildahsv}) we arrive at,
\begin{equation}\label{vprim}
\tilde{V}'(r)\equiv \frac{R^4}{r_F^{C+4}}[(Ar^{A-1}-\frac{Br^{B-1}}{r^{B-A}_{h}})\cosh ^2 \eta- Cr^{C-1}\sinh ^2 \eta],
\end{equation}
\begin{equation}\label{vpprim}
\tilde{V}''(r)\equiv \frac{R^4}{r_F^{C+4}}[(A(A-1)r^{A-2}-\frac{B(B-1)r^{B-2}}{r^{B-A}_{h}})\cosh ^2 \eta- C(C-1)r^{C-2}\sinh ^2 \eta].
\end{equation}
At the next step from equation (\ref{imv2}),  we are going to find  Im$V_{Q\bar{Q}}$  and  then  applying  negative condition on the corresponding potential. So the subsequent relation should be satisfied by the following expression,
\begin{eqnarray}\label{cond}
\frac{(Ar_c^{A-1}-\frac{Br_c^{B-1}}{r^{B-A}_{h}})\cosh ^2 \eta- Cr_c^{C-1}\sinh ^2 \eta}{2((A(A-1)r_c^{A-2}-\frac{B(B-1)r_c^{B-2}}{r^{B-A}_{h}})\cosh ^2 \eta- C(C-1)r_c^{C-2}\sinh ^2 \eta)} >\nonumber\\ \frac{(r_c^A-\frac{r_c^B}{r^{B-A}_{h}})\cosh ^2 \eta- r_c^C\sinh ^2 \eta}{(Ar_c^{A-1}-\frac{Br_c^{B-1}}{r^{B-A}_{h}})\cosh ^2 \eta- Cr_c^{C-1}\sinh ^2 \eta},
\end{eqnarray}
 Here we note that,  due to the equation (\ref{imv2}) ,  (\ref{mtildahsv})  and the result of $\tilde{M}(r_{c}) $ one  can  find,
\begin{equation}\label{condtanh}
\tanh^{2}\eta < \frac{1}{r_c^{C-A}}-\frac{1}{r_c^{C-A}}(\frac{r_c}{r_h})^{B-A}.
\end{equation}
It is clear  that $\tanh^{2}\eta$ is always non-negative. On the other hand, we know that $r_c >r_h$, as above result and  the validity of equation (\ref{condtanh}) we need to have,
\begin{equation}\label{ba}
B-A <0.
\end{equation}
 Since we are working in the near horizon limit,   the values of $r_c-r_h$  will be  small,  so we have,
\begin{equation}\label{tanhcond}
\tanh \eta < \frac{1}{r_h^{\frac{C-A}{2}}} \varepsilon,
\end{equation}
where $\varepsilon$ is  small . Therefor there are two possibilities in here, one is the $C\geq A$ which leads to a very small rapidity ($\eta\leq\varepsilon$) and second one is  $C <A$, in this case if the coefficient of the $\varepsilon$ in the right hand side of above expression is large enough, $\eta$ can be larger than last case. In other words, for $z<1$ , the $\eta$ is  negligible but for $z\geq 1$ , the  $\eta$  is small.\\
 By considering the near horizon limit in equation (\ref{cond}) limit  $r_c\longrightarrow r_h$, we proceed the following result,
\begin{eqnarray}\label{cond2}
\frac{(A-B)r_h^{A-1}\cosh ^2 \eta- Cr_h^{C-1}\sinh ^2 \eta}{2(((A(A-1)-B(B-1))r_{h}^{A-2})\cosh ^2 \eta- C(C-1)r_{h}^{C-2}\sinh ^2 \eta)} >\nonumber\\
 \frac{- r_{h}^C\sinh ^2 \eta}{(A-B) r_h^{A-1}\cosh ^2 \eta- Cr_{h}^{C-1}\sinh ^2 \eta}.
\end{eqnarray}
Now, it is clear that for $z\geq 1$  the $\eta$ will be negligible,  so the result in equation (\ref{cond2})  with the limit of $\eta\longrightarrow 0$ will reduces to the following expression,
\begin{equation}\label{const}
A+B-1>0
\end{equation}
But, if we consider  the general values of velocity then the equation (\ref{cond}) will be as,
\begin{equation}\label{const2}
(A-B)^2 X^2+2(A-B)(A+B-C-1) X+C(2-C)>0,
\end{equation}
where $X=r_h^{A-C} \coth^{2}\eta$.
If we take (\ref{const2}) as $aX^2+bX+c$, we can see  $\Delta=b^2-4ac$ is always positive. So there are two roots that the expression between them is negative and out of this region the expression is positive.
It means $X$ with values less than the first root or with values more than second root make the expression positive. Since we want to find some exact constraint according to A,B and C to satisfy (\ref{imv2}), we can not have analytical result for cases with non-small $\eta$.
Finally, one can say that for $z\geq 1$ the $\eta$ is very small for the validity of imaginary part of potential. On the other hand, in order to have exact solution according to $z$, $d$ and $\theta$ for the case of $z\leq 1$ the $\eta$ must be small. In that case we have also the validity of imaginary part of corresponding potential. We note here that case of $\theta=4$ with $z=1$  and also $\theta=3$ with $z=0$ by considering $d<3$ from equation (\ref{solution}), satisfy constraint (\ref{ba}). Such obtained dynamical parameter in hyperscaling metric background agree with the result of papers \cite{tsc,hybb,jqp}.

\section{Validity of Im$V_{Q\bar{Q}}$ for  dipole at arbitrary angles}
 After considering perpendicular case in the previous section, in this section we extend our calculations for arbitrary angles. It means that the orientation of dipole can have any arbitrary angle with respect to the velocity . As before we are supposed to extract the imaginary parts of potential by the method of paper\cite{sif} . $ \Theta$ is the angle of the dipole with respect to the $ X_{d-1}$ and dipole is placed on the $ (X_1,X_{d-1})$  plane. We have two degrees of freedom $ r(\sigma)$ and $ X_{d-1}(\sigma)$ for the arranging partition function of string, so the  string partition function will be following form,
\begin{equation}
\label{eq:thermalflucpartang}
Z_{c} \sim \int D(\delta r) \, D(\delta X_{d-1}) e^{i S_{c} (\bar{r}+\delta r, \bar{X}_{d-1} + \delta X_d)},
\end{equation}
where fluctuations $\delta r(\sigma)$ and $ \delta X_{d-1}(\sigma)$ are considered with $ \frac{\partial r}{\partial\sigma}\longrightarrow 0$ and $ \frac{\partial X_{d-1}}{\partial\sigma}\longrightarrow 0$.\\
By using the action of string and considering the interval divided to  $ 2N$ subintervals we  will arrive at,
\begin{equation}\label{partfuncarbteta}
Z_{c} \sim \left( \int_{-\infty}^{\infty} d(\delta r_{-N}) \, d(\delta X_{{d-1},-N}) \right) \cdots \left( \int_{-\infty}^{\infty} d(\delta r_{N}) \, d(\delta X_{{d-1},N}) \right) e^{i \frac{\mathcal{T} \Delta x}{2\pi \alpha'} \mathcal{L}_j},
\end{equation}
and
\begin{equation}\label{lagrangy}
\mathcal{L}_j = \sqrt{\tilde{M}(r(x_j)) (r'(x_j))^2 + V(r(x_j)) (X_{d-1}'(x_j))^2 + \tilde{V} (x_j)}.
\end{equation}
We expand the classical solution $ \bar{r}(0)$ around $\sigma=0$ to quadratic order on $\sigma$. If the string did not sag, then we would have $ X_{d-1}(\sigma)=\frac{\sigma}{\tan \tilde {\theta}}$ .  Also the $X_d(\sigma)$ must be an odd function of $ \sigma$  under reflections with respect to the the $(X_1,X_d)$ plane. So, in that case the  $X_{d-1}(\sigma)$ around $\sigma=0$ will be following form,
\begin{equation}\label{xd1}
X_{d-1}(\sigma) = \frac{\sigma}{\tan \tilde{\theta}} + b \sigma^3 + O (\sigma^5),
\end{equation}
where $\tilde{\theta}$ is equal to $\Theta$ and $b$ is a constant. So, from equation (\ref{xd1}) one can obtain,
\begin{equation}\label{xd11}
X_{d-1}'(\sigma)^2 = \frac{1}{\tan^2 \tilde{\theta}} + \frac{6 b}{\tan \tilde{\theta}} \sigma^2.
\end{equation}
The substitution of equation (\ref{xd11}) into  (\ref{lagrangy}), one can arrive at,
\begin{equation}\label{lagrangy2}
\mathcal{L}_j = \sqrt{\tilde{C}_1 x_j^2 + \tilde{C}_2},
\end{equation}
where $\tilde{C}_1$ and $\tilde{C}_2$ are respectively,
\begin{equation}\label{c1arbteta}
\tilde{C}_1 \equiv \tilde{M}_c\bar{r}''(0)^2+ \frac{1}{2} \left(\frac{V'_c}{\tan^2 \tilde{\theta}}+\tilde{V}'_c\right)\bar{r}''(0)+\frac{6 b}{\tan \tilde{\theta}} V_c
\end{equation}
\begin{equation}\label{c2arbteta}
\tilde{C}_2 \equiv  \left(\frac{V_c}{\tan^2 \tilde{\theta}}+\tilde{V}_c\right) + \left(\frac{V'_c}{\tan^2 \tilde{\theta}}+\tilde{V'}_c\right) \delta r + \left(\frac{V''_c}{\tan^2 \tilde{\theta}}+\tilde{V''}_c\right) \frac{(\delta r)^2}{2}.
\end{equation}
Also here we do  some algebraic calculations  same as before and obtain the $ImV_{Q\bar{Q}}$,
\begin{equation}
\label{eq:ImFQQang}
\mathrm{Im}\,V_{Q\bar{Q}} = -\frac{1}{4\alpha'}\frac{1}{\sqrt{\tilde{C}_1}} \left[ \frac{\left(\frac{V_c'}{\tan^2\tilde{\theta}}+ \tilde{V}_c'\right)^2}{2 \left(\frac{V_c''}{\tan^2\tilde{\theta}}+ \tilde{V}_c''\right)} - \left(\frac{V_c}{\tan^2\tilde{\theta}}+ \tilde{V}_c\right) \right]\,.
\end{equation}
In order to calculate the bracket of equation (\ref{eq:ImFQQang}), we use equations of (\ref{g00}), (\ref{gxx}), (\ref{grr}) and (\ref{vtildahsv}).  So, the expression of bracket (\ref{eq:ImFQQang}) in near horizon limit will be as,
\begin{eqnarray}\label{braket}
[.....]&=&\frac{R^4}{r_F^{C+4}}[\frac{((A-B)r_h^{A-1}(\frac{1}{\tan^2\tilde{\theta}}+\cosh^2\eta)-Cr_h^{C-1}\sinh^2\eta)^2}{2(r_h^{A-2}(A(A-1)-B(B-1))(\frac{1}{\tan^2\tilde{\theta}}+ \cosh^2\eta)-C(C-1)r_h^{C-2}\sinh^2\eta   )} \nonumber\\
&-&((A-B) r_h^{A-1}\cosh^2\eta -Cr_h^{C-1}\sinh^2\eta)].
\end{eqnarray}
Here the condition of Im$V_{Q\bar{Q}}<0$ lead us to consider,
\begin{eqnarray}\label{inegativ}
& &(2[(A-B)(A+B-1)-C(C-1)r_h^{C-A}][A-B-Cr_h^{C-A}]-r_h(A-B-Cr_h^{C-A})^2) {\sinh^4\eta} \nonumber\\
 &+&(2[(A-B)^2(A+B-1)-C(C-1)(A-B) r_h^{C-A}+\frac{(A-B)(A+B-1)(A-B-Cr_h^{C-A})}{\cos^2\tilde{\theta}}]\nonumber\\
 &-&2r_h(\frac{A-B}{\cos^2\tilde{\theta}})(A-B-Cr_h^{C-A})) {\sinh^2\eta}  \nonumber\\
&+&(2\frac{(A-B)^2(A+B-1)}{\cos^2\tilde{\theta}} -r_h(\frac{A-B}{\cos^2\tilde{\theta}})^2)<0.
\end{eqnarray}
The small values of $\eta$ (\ref{inegativ})  and validity of condition on the corresponding potential give us the following constraint,
\begin{equation}\label{constarbteta}
2(A+B-1) < \frac{r_h}{\cos^2\tilde{\theta}}
\end{equation}
On the other hand  in the near horizon for the case of  $\tilde{C}_1 >0$  with the small values of $\eta$ from(\ref{c1arbteta}),  one can arrive at,
\begin{equation}\label{constrarb}
r_h \bar{r}''(0)>\frac{B-A}{2 \tan^2\tilde{\theta}}.
\end{equation}

 Here we can say that the relation (\ref{constrarb}) depends on the conditions of the problem. Some condition  should be taken into account with the numerical value of $\bar{r}''(0)$ and other parameters. In that case the equation (\ref{constrarb}) by using some suitable condition help us to obtain the definite corresponding potential and arrange the dynamic parameter and dimension in hyperscaling violation metric background. Anyway  we can  not find an analytic expression that assures the region where the imaginary part of potential is negative. For this reason we leave this problem for the future.  But there is some good news for the equation (\ref{constarbteta}).  We  can compare two important  cases from previous section and this section as perpendicular and arbitrary angle respectively. The common conditions for two cases can be arrange by the parametres  $\theta$, $z$ and $d$ which are  satisfied by Im$V_{Q\bar{Q}}<0$ in perpendicular and arbitrary angle. For two cases the corresponding parameters will be as $\theta=4$, $z=1$, $d=1or 2$ and  $\theta=3$, $z=0$, $d=1or 2$. By using (\ref{constarbteta}), the first one gives these results for  $d=1$ and   $d=2$
\begin{eqnarray}
\cos^2\tilde{\theta} <\frac{1}{42} r_h, \nonumber\\
\cos^2\tilde{\theta} <\frac{1}{12} r_h.
\end{eqnarray}
and the other one gives us,
\begin{eqnarray}
 \cos^2\tilde{\theta} <\frac{1}{34} r_h, \nonumber\\
 \cos^2\tilde{\theta} <\frac{1}{12} r_h.
\end{eqnarray}
 Where they show that a larger spatial dimension leads to a larger maximum value of $\cos^2\tilde{\theta}$ or smaller angle. For the $d=2$,  both cases have the same limit of value of arbitrary angle which is interesting  for the some phenomena in QCD.\\
Proceeding by considering imaginary part of potential for corresponding cases, we use these condidate parameters  to study behavior of a moving meson in a plasma with hyperscaling violation metric background. With respect to what we found previously, we study following graphs. Notice here that we fix an arbitrary angle and use one of these mentioned category of $\theta$, $z$ and $d$ which we saw satisfy conditions.
\begin{figure}[h!]
\begin{center}$
\begin{array}{cccc}
\includegraphics[width=100 mm]{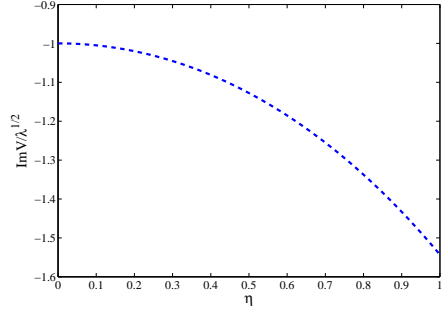}
\end{array}$
\end{center}
\caption{Imaginary part of potential of a moving meson in a plasma with respect to rapidity of that, with acceptable values of hyperscaling violation parameters and at a fixed arbitrary angle.}
\label{fig:imveta}
\end{figure}
\begin{figure}[h!]
\begin{center}$
\begin{array}{cccc}
\includegraphics[width=100 mm]{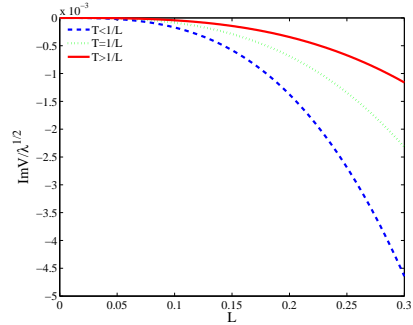}
\end{array}$
\end{center}
\caption{Imaginary part of potential of a moving meson in a plasma as a function of $L$ (inter-quark distance length) in different temperatures, with acceptable values of hyperscaling violation parameters and at a fixed arbitrary angle and rapidity.}
\label{fig:imvl}
\end{figure}

Figure (\ref{fig:imveta}) shows the imaginary part of potential with respect to the rapidity of meson in plasma where one class of acceptable $d$, $\theta$ and $z$ have been taken into account. One can see the magnitude of ImV increases with increasing rapidity. This result is in agreement with QGP results based on the NLO perturbative QCD calculation, show that the thermal width becomes larger as the quarkonium travels faster with respect to the QGP \cite{teasoo}. Also we notice that our results are presented for small rapidity in a plasma with hyperscaling violation background and this fact arose from condition $ImV<0$. It is comparable with studying the dissociation machanism by means of in QCD side \cite{hqm}, where at some moderate velocities the width is an increasing function of velocity. In figure (\ref{fig:imvl}) we consider behavior of ImV with respect to inter-quark distance length in different temperatures where the temperature has been defined in relation (\ref{metric4}). Studying the graph, one can see for increasing temperature, the onset of the imaginary part happens for larger $L$.  Therefore, while with
decreasing temperature the imaginary part of the potential starts to become nonzero for smaller
values of $L$, its magnitude considerably diminishes with increasing temperature. This result is comparable with QCD case which has been studied in \cite{hqm}, it shows our result at small velocity in HSV background are related to QCD cases at moderate velocities, such as calculations of s-wave states and NLO dissociation cross section in high energy limit \cite{teasoo}.\\
 Calculations of ImV relevant to QCD and heavy ion collision were performed for static $Q\bar{Q}$ pair using perturbative QCD (pQCD) in \cite{static}. They used classical approximation to study ImV  as we did. In standard calculation of ImV for $N=4$SYM \cite{sif} ImV roughly scales as $T^2$, opposed  to the $T$ scaling predicted by pQCD. In our work and with mentioned conditions ImV scales as $T^{(\frac{2\theta}{d}-z)}$ (absoloutly $\eta$ should be small in this scaling). In compare with pQCD, we can study our work. In both cases potential of a pair has an imaginary part which implys the thermal effects generate a decay width. In exact static case for example in decay width of bottomonium system, one can see an increasing behavior as a function of $T$ while in our slowly moving case with a HSV gravity dual, the magnitude of ImV decreases with temperature as it happens for some hot QCD cases as it has been explained before. In \cite{static} thermal width can be ignored for $T\leq g^2\frac{M}{12\pi}$ where $M$ is heavy mass and $g^2$ is strong gauge coupling. But in our results the magnitude of thermal width becomes smaller in higher temperatures. So, phenomenologically there is an upper bound of temperature which can be found numerically for any required HSV background where ImV gets zero value as  \cite{static} introduced the lower bound of temperature (a temperature before that ImV is ignored, so we interpreted it as lower bound) in QCD side of duality.\\
Recently, the effect of finite 't Hooft coupling correction on the imaginary potential of a moving heavy quarkonium from the AdS/CFT has been studied \cite{iptw}. As a comparison we consider ImV depends on velocity absolutly in both our and their works. In mentioned work the imaginary potential is defined in the region $(L_{min}, L_{max})$ and in our holography approach, $L_{max}$ can exactly be defined and one should find numerically in corresponding problem. The authors in \cite{iptw} insist  only slowly moving quarkonium can be considered, because for large velocities the valid region of the saddle point approximation strongly decreases and the absolute value of the imaginary part of potential vanishes. This is comparable with what we saw in calculation of ImV with HSV background where the  condition $ImV<0$  imposes to have just small velocities. In addition in their approximate soultion of thermal width, graph shows an increasing behaviour versus the rapidity which is in agreement with our results. The two different approaches of that paper are related to small velocities and they can be found in a weekly coupled plasma \cite{teasoo,hqm}. In the approximate solution for a moving heavy quarkonium \cite{ihqm} it is found that increasing of velocity leads to higher decay rate for the quarkonium, which is in agreement with our result where we considered in figure(\ref{fig:imveta}) that magnitude of ImV increases with increasing repidity.
\section{Concolusion}
In this paper  we  checked the imaginary part  of corresponding potential for the  moving quarkonia in plasma with Lifshitz-like backgrounds which are known as hyperscaling violation metrics. So, at the first step we considered some string solutions  due  to Im$V_{Q\bar{Q}}<0$  and found  some  good results which they  constrained $d$, $\theta$ and $z$ with suitable values. Since a moving quarkonia in plasma can have  two cases as  perpendicular velocity vector with respect to joining axis of the pair or an arbitrary angle, we  applied our process for   two parts and found  some conditions or constraints. All obtained conditions for the $\eta$, $d$, $z$ and $\theta$ were satisfied by Im$V_{Q\bar{Q}}<0$ for two cases in same time.
As we know  people have been found many string solutions for the hyperscaling violation metrics at the near horizon limit over the last few years.  In order to have Im$V_{Q\bar{Q}}<0$ we used the above limit and  we found  for $ z\geq 1$ just small values of  rapidity $\eta$ must  be taken into account, otherwise for $z <1$, $\eta$ can be small. Here we have shown that our results strongly depend on $\eta$, such parameter implies that gravity duals are related to some QCD phenomena. So, one important result here is that these  two  values of $z$ lead to different behavior of the moving quarkonia in plasma. But the results shown  that  there is no analytic constraint according to the case when larger values of $\eta$ is considered. Therefore, it completely depends on the conditions of the case. In this paper , we applied the first one with constant dilaton $z=0$ and the other one  dynamical dilaton, $z=1$. For  two cases   we found the suitable values for $d$ to be satisfied by imaginary part of the potential for a YM theory in finite temperature. Therefore, the spatial dimensions of these two cases  are coincident. Also, for arbitrary angles section we followed all the steps as before.
So, at near horizon limit for very small values of $\eta$ we found two analytical and numerical constraints. Depending on the  corresponding problem, one can try to use analytical expression and the other should be coincident with that numerically. Analytic conditions gave us uper limit of $\cos^2\tilde{\theta}$ for two corresponding cases. Briefly, we have seen  that the case  $\theta=3$, $z=0$, $d=1 or  2$ with constant dilaton is similar to $\theta=4$, $z=1$, $d=1 or  2$ with dynamical dilaton for this respect to the spatial dimension and arbitrary angles. Thus  in order to have Im$V_{Q\bar{Q}}<0$  we have shown that the hyperscaling violating metric is very close to a YM theory with constant dilaton and finite temperature with suitable $d$, $z$, $\theta$ and $\eta$.\\
Generally in the context of decay width, the quarkonium mesons are stable for $T<T_{diss}$ and disappear for $T>T_{diss}$. As we discuss in introduction, a Yang-Mills theory at $T\neq 0$ with QCD at $T>T_{c}$ have some similarities. As we discussed on ImV at nonzero temperature. Our results indicate that magnitude of ImV decreases with increasing temperature which is comparable with hot QCD case studied in \cite{hqm}. We saw condition $ImV<0$ imposes that magnitude of thermal width becomes smaller in higher temperature. Therefore, phenomenologically it gives us an upper bound of temperature which can be found numerically for any  HSV gravity dual of corresponding QCD. Briefly,  the temperature in that ImV dissapears, introduces the dissociation  temperatures and binding energies of the different quarkonia states. It is worth to notify this is in result that the overal potential is negative.\\

{\bf Acknowledgement}\\
The authors are grateful very much to Fatemeh Razavi for support and valuable activity in numerical calculations.


\begin{thebibliography}{99}
\bibitem{jmm} J. M. Maldacena, \emph{The Large N Limit of Superconformal Field Theories and Supergravity}, \emph{ Adv.Theor.Math.Phys}, {\bf 2} (1998) 231 [arXiv: 9711200 [hep-th]] .
\bibitem{ssg} S. S. Gubster, I. R. Klebanov and A. M. Polyakov, \emph{Gauge Theory Correlators from Non-Critical String Theory}, \emph{Phys. Lett. B}, {\bf 428} (1998) 105 [arXiv: 9802109 [hep-th]].
\bibitem{ew} E. Witten, \emph{Anti De Sitter Space And Holography}, \emph{Adv. Theor. Math. Phys}, {\bf 2} (1998) 253 [arXiv: 9802150 [hep-th]]
\bibitem{oas} O. Aharony, S. S. Gubster, J. M. Maldacena, H. Ooguri and Y. Oz, \emph{Large N Field Theories, String Theory and Gravity}\emph{Phys. Rept}, {\bf 323}  (2000) 183 [arXiv: 9905111 [hep-th]]
\bibitem{kli}Klebanov, Igor R. et al, \emph{AdS / CFT correspondence and symmetry breaking} \emph{Nucl.Phys.B} {\bf 556}(1999)89 [arXiv:9905104[hep-th]]
\bibitem{bim} Bianchi, Massimo et al, \emph{Holographic renormalization} \emph{Nucl.Phys. B} {\bf 631}(2002)159 [arXiv:0112119 [hep-th]]
\bibitem{darl}Da Rold, Leandro et al , \emph{Chiral symmetry breaking from five dimensional spaces} {\emph Nucl.Phys.B} {\bf 721}(2005) [arXiv:0501218 [hep-ph]]
\bibitem{otpt}'t Hooft, Gerard \emph{On the Phase Transition Towards Permanent Quark Confinement} \emph{Nucl.Phys. B} {\bf 138}(1978)1
\bibitem{crbe}Svetitsky, Benjamin et al,     \emph{Critical Behavior at Finite Temperature Confinement Transitions} \emph{Nucl.Phys. B} {\bf 210}(1982)423;
\bibitem{hoba} M. R. Pahlavani, J. Sadeghi and R. Morad \emph{The holographic description of a baryon in non-critical string theory} \emph{J. Phys. G; Nucl. Part. Phys} {\bf 39} (2012) 065004.
\bibitem{nbma} N. Brambilla, M. A. Escobedo, J. Ghiglieri, J. Soto and A. Vairo, \emph{Heavy Quarkonium in a weakly-coupled quark-gluon plasma below the melting temperature}, \emph{JHEP}, {\bf 09} (2010) 038 [arXiv:1007.4156 [hep-ph]].
\bibitem{ybm} Y. Guo and M. Strickland, \emph{The imaginary part of the static gluon propagator in an anisotropic (viscous) QCD plasma},\emph{Phys. Rev. D}, {\bf 79} (2009) 114003 [arXiv:0903.4703[hep-ph]].
\bibitem{mlop} M. Laine, O. Philipsen, P. Romatschke and M. Tassler, \emph{Real-time static potential in hot QCD}, \emph{JHEP}, {\bf 0703} (2007) 054 [arXiv:0611300 [hep-ph]].
\bibitem{arth} A. Rothkopf, T. Hatsuda and S. Sasaki, \emph{Complex Heavy-Quark Potential at Finite Temperature from Lattice QCD}, \emph{Phys. Rev. Lett}. {\bf 108} (2012) 162001 [arXiv:1108.1579 [hep-lat]].
\bibitem{gaca} G. Aarts, C. Allton, S. Kim, M. P. Lombardo, M. B. Oktay, S. M. Ryan, D. K. Sinclair and J. I. Skullerud, \emph{What happens to the Upsilon and $eta_b$ in the quark-gluon plasma? Bottomonium spectral functions from lattice QCD}, \emph{JHEP},  {\bf 1111}  (2011)  103  [arXiv:1109.4496[hep-lat]].
\bibitem{gcs} G. Aarts, C. Allton, S. Kim, M. P. Lombardo, S. M. Ryan and J. I. Skullerud, \emph{Melting of P wave bottomonium states in the quark-gluon plasma from lattice NRQCD}, \emph{JHEP},  {\bf 1312}  (2013) 064 [arXiv:1310.5467 [hep-lat]].
\bibitem {astp}E. Witten, \emph{Anti-de Sitter space, thermal phase transition, and confinement in gauge
theories}, \emph{Adv. Theor. Math. Phys}, {\bf 2} (1998) 505, [arXiv:9803131 [hep-th]].
\bibitem{adsqcd} J. Casalderrey-Solana, H. Liu, D. Mateos, K. Rajagopal and U. Achim Wiedemann, \emph{Gauge/String Duality, Hot QCD and Heavy Ion Collisions}, \emph{CERN-PH-TH/2010-316, MIT-CTP-4198, ICCUB-10-202}, [arxiv: 1101.0618 [hep-th]].
\bibitem{ddsc}  M.- C. Chu and T. Matsui, \emph{Dynamic Debye screening for a heavy-quark-antiquark pair traversing a quark-gluon plasma}, \emph{Phys. Rev. D} {\bf 39}  (1989) 1892.
\bibitem{mst} M. Strickland,\emph{Thermal Upsilon(1s) and $chi_b1$ suppression in $sqrt(s_NN)=2.76 TeV$ Pb-Pb collisions at the LHC},\emph{ Phys. Rev. Lett} {\bf 107}  (2011) 132301 [arXiv:1106.2571[hep-ph]].
\bibitem{msd} M. Strickland and D. Bazow,\emph{Thermal Bottomonium Suppression at RHIC and LHC}, \emph{ Nucl. Phys. A} {\bf 879}  (2012)  25 [arXiv:1112.2761[nucl-th]].
\bibitem{mmk} M. Margotta, K. McCarty, C. McGahan, M. Strickland, and D. Yager-Elorriaga, \emph{Quarkonium states in a complex-valued potential}, \emph{Phys.Rev.D} {\bf 83}   (2011) 105019 [arXiv:1101.4651 [hep-ph]].
\bibitem{gac} G.Aarts, C. Allton, S. Kim, M. P. Lombardo, M. B. Oktay, S. M. Ryan, D. K. Sinclair and J. I. Skullerud,\emph{S wave bottomonium states moving in a quark-gluon plasma from lattice NRQCD}, \emph{ JHEP}  {\bf 1303}  (2013) 084 [arXiv:1210.2903[hep-lat]].
\bibitem{nbs} M. A. Escobedo, M. Mannarelli and J. Soto , \emph{Non-relativistic bound states in a moving thermal bath} \emph{Phys.Rev.D} {\bf 84}  (2011) 016008  [arXiv:1105.1249v2[hep-ph]].

\bibitem{drst} J. Sadeghi, M. R. Setare \emph{Drag force with different charges in STU background and AdS/CFT}, \emph{J. Phys.G; Nucl.Part. Phys} {\bf 36} (2009)115005.
\bibitem{sif} S. I. Finazzo and J. Noronha,\emph{Thermal suppression of moving heavy quark pairs in strongly coupled plasma} [arXiv: 1406.2683[hep-th]].
\bibitem{tec} J. Sadeghi and S. Tahery , \emph{The effects of deformation parameter on thermal width of moving quarkonia in plasma}, \emph{JHEP} {\bf 06}  (2015)  204 [arXiv:1412.8332v3 [hep-th]].
\bibitem{ipap} K. B. Fadafan, D. Giataganas and H. Soltanpanahi,  \emph{The Imaginary Part of the Static Potential in Strongly Coupled Anisotropic Plasma}, \emph{JHEP} {\bf 11 } (2013) 107  [arXiv:1306.2929 [hep-ph]].
\bibitem{osca} D. Giataganas,  \emph{Observables in Strongly Coupled Anisotropic Theories}, \emph{ Contribution to the Proceedings of the XVIII European Workshop on String Theory}  (2012)  [arXiv: 1306.1404v3[hep-th]].
\bibitem{ihqm} M. Ali-Akbari, D. Giataganas and Z. Rezaei, \emph{The Imaginary Potential of Heavy Quarkonia Moving in Strongly Coupled Plasma}, \emph{Phys. Rev. D} {\bf 90 } (2014) 086001  [arXiv:1406.1994 [hep-th]].
\bibitem{fn} S. I. Finazzo and J. Noronha,\emph{Estimates for the Thermal Width of Heavy Quarkonia in Strongly Coupled Plasmas from Holography}, \emph{JHEP}  {\bf 1311}   (2013)  042 [arXiv:1306.2613[hep-ph]].
\bibitem{pd} P. Dey and S. Roy, \emph{Lifshitz-like space-time from intersecting
space-time},\emph{ Phys. Rev. D}, {\bf 86}  (2012) 066009  [arXiv:1204.4858[hep-th]].
\bibitem{sr} P. Dey and S. Roy, \emph{Lifshitz-like space-time from intersecting
branes in string/M theory},\emph{ JHEP}, {\bf 06}  (2012)  129 [arXiv:1203.5381[hep-th]].
\bibitem{tsc} J. Sadeghi, B. Pourhassan, F. Pourasadollah, \emph{Thermodynamics of Schrödinger black holes with hyperscaling violation} , \emph{ 	Phys. Lett. B} , {\bf 720 } (2013)244 [arXiv:1209.1874 [hep-th]]
\bibitem{hybb} J. Sadeghi, A. Asadi, \emph{Hydrodynamics in black brane with hyperscaling violation metric background}, \emph{Can.J.Phys}, {\bf92 } (2014) 1570 [arXiv: 1404.5282[hep-th]]
\bibitem{jqp} J. Sadeghi, S. Heshmatian, \emph{Jet Quenching Parameter with Hyperscaling Violation} , \emph{Eur.Phys.J. C}, {\bf74 } (2014)  3032[arXiv: 1308.5991[hep-th]]
\bibitem{jpol}J. Polchinski and M.J. Strassler, \emph{Hard Scattering and Gauge/String Duality},\emph{ Phys. Rev. Lett}, {\bf 88}, (2002) 031601. [arXiv:0109174[hep-th]]. 
\bibitem{ppc} S. S. Gubser and F. D. Rocha, \emph{Peculiar properties of a charged dilatonic black hole in
AdS5},\emph{ Phys. Rev. D}, {\bf 81}  (2010)  046001 [arXiv:0911.2898 [hep-th]].
\bibitem{hoc}  K. Goldstein, S. Kachru, S. Prakash and S. P. Trivedi, \emph{Holography of Charged
Dilaton Black Holes},\emph{JHEP }, {\bf 1008}  (2010)  078 [arXiv: 0911.3586[hep-th]].
\bibitem{ptb} M. Cadoni, G. D'Appollonio and P. Pani, \emph{Phase transitions between
Reissner-Nordstrom and dilatonic black holes in 4D AdS spacetime},\emph{ JHEP}, {\bf 1003}  (2010) 100  [arXiv: 0912.3520[hep-th]].
\bibitem{hcd} M. Cadoni and P. Pani, \emph{Holography of charged dilatonic black branes at finite
temperature},\emph{JHEP }, {\bf 1104}  (2011) 49  [arXiv: 1102.3820[hep-th]].
\bibitem{eff} C. Charmousis, B. Gouteraux, B. S. Kim, E. Kiritsis and R. Meyer, \emph{Effective
Holographic Theories for low-temperature condensed matter systems}, \emph{JHEP }, {\bf 1011}  (2010) 151   [arXiv:1005.4690[hep-th]].
\bibitem{dwh} E. Perlmutter, \emph{Domain Wall Holography for Finite Temperature Scaling Solutions}, \emph{JHEP }, {\bf 1102}  (2011) 013   [arXiv:1006.2124[hep-th]].
\bibitem{tbc} G. Bertoldi, B. A. Burrington and A. W. Peet, \emph{Thermal behavior of charged dilatonic
black branes in AdS and UV completions of Lifshitz-like geometries}, \emph{ Phys. Rev. D}, {\bf 82}  (2010) 106013  [arXiv: 1007.1464[hep-th]].
\bibitem{hddb} K. Goldstein, N. Iizuka, S. Kachru, S. Prakash, S. P. Trivedi and A. Westphal, \emph{Holography of Dyonic Dilaton Black Branes}, \emph{JHEP }, {\bf 1010}  (2010)  027 [arXiv:1007.2490[hep-th]].
\bibitem{hfn} N. Iizuka, N. Kundu, P. Narayan and S. P. Trivedi, \emph{Holographic Fermi and
Non-Fermi Liquids with Transitions in Dilaton Gravity},  [arXiv:1105.1162[hep-th]].
\bibitem{cdab} P. Berglund, J. Bhattacharyya and D. Mattingly,  \emph{Charged Dilatonic AdS Black
Branes in Arbitrary Dimensions},   [arXiv:1107.3096[hep-th]].
\bibitem{hfs} N. Ogawa, T. Takayanagi and T. Ugajin, \emph{Holographic Fermi Surfaces and
Entanglement Entropy},  [arXiv:1111.1023[hep-th]].
\bibitem{hfsi} L. Huijse, S. Sachdev and B. Swingle, \emph{Hidden Fermi surfaces in compressible states of
gauge-gravity duality},  [arXiv:1112.0573[cond-mat.str-el]]
\bibitem{heef} E. Shaghoulian, \emph{Holographic Entanglement Entropy and Fermi Surfaces}  [arXiv:1112.2702[hep-th]].
\bibitem{scsd} D. S. Fisher, \emph{Scaling and critical slowing down in random field Ising systems}, \emph{Phys.
Rev. Lett. }, {\bf 56}  (1986) 416.
\bibitem{hfnf} N. Iizuka, N. Kundu, P. Narayan and S. P. Trivedi, \emph{Holographic Fermi and
Non-Fermi Liquids with Transitions in Dilaton Gravity},  [arXiv:1105.1162[hep-th]].
\bibitem{ucbe} B. Swingle and T. Senthil, \emph{Universal crossovers between entanglement entropy and
thermal entropy}, [arXiv:1112.1069[hep-th]].
\bibitem{isfh} P. Dey and S. Roy \emph{Interpolating solution from AdS$_5$ to hyperscaling violating Lifshitz space-time},\emph{Phys. Rev. D }, {\bf 91}  (2014) 026005  [arXiv:1406.5992[hep-th]].
\bibitem{newhad} M. Laine, \emph{News on hadrons in a hot medium}, [arXiv:1108.5965 [hep-ph]].
\bibitem{qgpf} T. Matsui, H. Satz, \emph{J/psi Suppression by Quark-Gluon Plasma Formation}   \emph{Phys. Lett. B}  {\bf 178}   (1986 ) 416.
\bibitem{recrde}  M. A. Escobedo, \emph{The relation between cross-section, decay width and imaginary potential
of heavy quarkonium in a quark-gluon plasma,}   \emph{J. Phys. Conf. Ser,}  {\bf 503} (2014) 012026  [arXiv: 1401.4892[hep-th]].
\bibitem{stf} N. Brambilla, J. Ghiglieri, A. Vairo, P. Petreczky,  \emph{Static quark−antiquark pairs at
finite temperature},  \emph{Phys. Rev. D}  {\bf 78 } (2008 )  014017 [arXiv:0804.0993 [hep-ph]].
\bibitem{hqwe} N. Brambilla, M. A. Escobedo, J. Ghiglieri, J. Soto, A. Vairo, \emph{Heavy Quarkonium
in a weakly-coupled quark-gluon plasma below the melting temperature}, \emph{JHEP}  {\bf 1009}  (2010 ) 038   [arXiv:1007.4156 [hep-ph]].
\bibitem{pNRQCD} N. Brambilla, M. A. Escobedo, J. Ghiglieri, A. Vairo, \emph{Thermal width and gluodissociation
of quarkonium in pNRQCD}   \emph{JHEP}  {\bf 1112} ( 2011) 116  [arXiv:1109.5826[hep-ph]]
 \bibitem{anis}  A. Dumitru, \emph{Quarkonium in a non-ideal hot QCD Plasma},   \emph{Prog. Theor. Phys. Suppl,}  {\bf 187} (2011) 87 [arXiv:1010.5218 [hep-ph]].
\bibitem{comhe} A. Rothkopf, T. Hatsuda, S. Sasaki,\emph{ Complex Heavy-Quark Potential at Finite Temperature
from Lattice QCD}, \emph{Phys. Rev. Lett}, {\bf 108} (2012) 162001 [arXiv:1108.1579[hep-lat]].
\bibitem{qstate} M. Margotta, K. McCarty, C. McGahan, M. Strickland, D. Yager-Elorriaga,  \emph{Quarkonium
states in a complex-valued potential.} \emph{ Phys. Rev. D}, {\bf 83}  (2011) 105019 [Erratumibid.D 84 (2011) 069902] [arXiv:1101.4651 [hep-ph]].
\bibitem{surv} B. Z. Kopeliovich, I. K. Potashnikova, I. Schmidt, M. Siddikov, \emph{Survival of charmonia
in a hot environment}, 	\emph{Phys. Rev. C}, {\bf 91},  (2015) 024911 [arXiv:1409.5147 [hep-ph]].
\bibitem{pbpb} F. Nendzig, G. Wolschin, \emph{ Bottomium suppression in PbPb collisions at LHC energies},
\emph{J. Phys. G}, {\bf 41} (2014) 095003 [arXiv:1406.5103 [hep-ph]].
\bibitem{teasoo} T. Song, et al \emph{The thermal width of heavy quarkonia moving in quark gluon plasma}, \emph{PLB}, {\bf 659} (2008) 621 [arXiv:0709.0794[hep-ph]].
\bibitem{hqm} M. A. Escobedo, F. Giannuzzi, M. Mannarelli and J. Soto , \emph{Heavy Quarkonium moving in a Quark-Gluon Plasma}, \emph{Phys.Rev.D} {\bf 87}  (2013)  114005 [arXiv:1304.4087 [hep-ph]]
\bibitem{wipo}Rey, Soo-Jong et al ,\emph{Wilson-Polyakov loop at finite temperature in large N gauge theory and anti-de Sitter supergravity}, \emph{Nucl.Phys. B}  {\bf527} (1998)191 [arXiv:9803135[hep-ph]]
\bibitem{ugkm} U. Gürsoy, E. Kiritsis, G. Michalogiorgakis and F. Nitti,\emph{Thermal Transport and Drag Force in Improved Holographic QCD}, \emph{JHEP} {\bf 0912} (2009) 056 [arXiv:0906.1890 [hep-ph]].
\bibitem{static} M. Laine et. al, \emph{Real-time static potential in hot QCD} , \emph{JHEP}, {\bf 03} (2007) 054 [arXiv: 0611300[hep-ph]].
\bibitem{iptw} K. B. Fadafan and S. K. Tabatabaei, \emph{The Imaginary Potential and Thermal Width of
Moving Quarkonium from Holography}, \emph{J. Phys. G}, {\bf 43} (2016) 9 [arXiv: 1501.00439[hep-th]].
\end{thebibliography}
\end{document}